\magnification  1200
\newbox\Ancha
\def\gros#1{{\setbox\Ancha=\hbox{$#1$}
   \kern-.025em\copy\Ancha\kern-\wd\Ancha
   \kern.05em\copy\Ancha\kern-\wd\Ancha
   \kern-.025em\raise.0433em\box\Ancha}}
\font\bigggfnt=cmr10 scaled \magstep 3
 2
\font\bigfnt=cmr10 scaled \magstep 1


\baselineskip 14.2 pt
\def\Par{\par\vskip 6 pt}
\def\eq#1{ \eqno({#1}) \qquad }
\vglue .5 in
\def\ie{{\it i.\ e.\ }}

\noindent{\bigggfnt  Non-unitary representations of the SU(2) algebra
 in the Dirac equation with a Coulomb potential} \Par

\vskip0.035truein
\vskip 6 pt

\noindent {\bigfnt R.\ P.\ Mart\'{\i}nez-y-Ro\-mero,\footnote{*}{ \rm On sabbatical leave from Facultad de Ciencias UNAM,\par e-mail: rodolfo@dirac.fciencias.unam.mx} A.\ L.\ Salas-Brito}\par

\noindent{\it Laboratorio de Sistemas Din\'amicos, Departamento de Ciencias B\'asicas, UAM-Az\-ca\-pot\-zalco. Apar\-tado Pos\-tal 21-726, Co\-yoa\-c\'an 04000 DF, M\'exico.\par
\noindent E-mail: asb@hp9000a1.uam.mx}\Par

\noindent {\bigfnt   Jaime Salda\~na-Vega }\par
\noindent {\it  Facultad de Ciencias, Universidad Nacional Aut\'onoma de M\'exico, \par 
\noindent Apartado Postal 50-542 M\'exico 04510 D.\ F.\ } \Par

\vskip 10 pt
\vskip 24pt
\vskip0.21truein

\centerline{\bigfnt Abstract.}\Par
\vskip0.035truein
\leftskip .3 in \rightskip .3 in \noindent   A novel realization of the classical SU(2) algebra is introduced for the Dirac relativistic hydrogen atom defining a set of operators that, besides, allow the factorization of the problem.   An extra phase is needed as a new variable in order to define the algebra. We take advantage of the operators  to solve the Dirac equation  using algebraic methods. To acomplish this, a similar  path to the one used in the angular momentum case is used; hence, the radial  eigenfuntions calculated  comprise  non unitary representations of the algebra. One of the interesting properties of such non unitary representations  is that they are not labeled by  integer nor by half-integer numbers as happens in the usual angular momentum representation.  \Par
\vskip 15pt

\vfil

\par

\eject

\noindent{\bf I. Introduction}\Par

 The unitary representations of groups and algebras are of a  great interest in physics. We
can  mention the ubiquitous example of  angular momentum or the SU(2)
algebra. In this instance, as it is well known, the representations are labeled
by two real numbers $j$ and $m$ which may take only integer or half-integer
values; the representations depend only on two parameters, $\theta$ and
$\phi$ say, which are defined in the  compact sets $[0,\,\pi]$ and
$[0,\,2 \pi]$ respectively.$^1$ However, there are many other physically interesting groups or algebras which are not unitary. In such a case, the representations are not  restricted to
parameters defined in compact sets nor its generators are  necessarily
Hermitian, but they nevertheless can play  an important role in physics. The Lorentz
group being a very important example of a physically relevant non compact group whose algebra is not necessarily unitary.$^2$ \Par

It is the purpose of this
article to introduce a realization of the cyclic $SU(2)$
algebra in terms of non Hermitian operators and then to use these operators  to factorize and solve  the relativistic Dirac hydrogen atom. The solution is obtained  using algebraic methods using the basic operators of the system, following a route parallel to the unitary compact case, but introducing an extra variable (found below to play the role of a phase) which is required in our approach.$^{3}$ \Par

 The use of algebraic techniques has been common for these sort of
problems. For example, in the relativistic hydrogen atom \`a la Dirac one approach has been the use of shift operators;$^4$ whereas in the non relativistic case a succesful approach requires the use of ladder operators and  the factorization method.$^{5\hbox{--}6}$ Our approach is more akin to the introduction of ladder operators than to the shift operator method customarily used for  this problem.$^4$\Par

The paper is organized as follows. In section  II we introduce the equations
of the problem and define  our notation. In section III we construct the basic operators spanning the SU(2) algebra as a useful tool for the problem. In section IV we define the inner product needed to  investigate the properties of our basic operators. In section V some non unitary representations  of the symmetry algebra are constructed as the radial eigenfunctions of the problem. We find that they are non compact representations  that play an equivalent role to the spherical harmonics in the unitary case. In order to write such representations we find convenient to define a family of polynomials that are found to be the associated generalized Laguerre polynomials.
In section VI, using algebraic methods and the operators defined in section II, we find the energy spectrum of the hydrogen atom.
In section VII we give our conclusions. In the Appendix, we list and plot the explicit expressions of the first 6  mentioned polynomials; they are  explicitly related  to the  generalized associated Laguerre polynomials previously used by Davis for expressing the radial eigenfunctions of the hydrogen atom.$^{7}$ \Par

\noindent{\bf II. The Dirac hydrogen atom}

The symmetry algebra for the bound states of the non-relativistic hydrogen atom, even in the classical case, is well known to be the $SO(4)$,$^{8,9}$ but in the relativistic case the situation is different as it is well known. The algebra associated with the symmetry of the radial part of the problem can be regarded as SU(2) as occurs with the angular momentum. The beautiful thing  is that we can then proceed to solve the relativistic hydrogen atom following a method that essentially parallels the calculation of eigenfunctions and eigenvalues of the angular momentum at the only  expense of introducing an extra phase.\Par

Let us begin with the Dirac Hamiltonian of the hydrogen atom

$$H_D = {\gros \alpha}\cdot\hbox{\bf p} + \beta m - {Ze^2\over
r}\,, \eq{1} $$

\noindent where $m$ is the mass  of an electron and ${\gros\alpha}$ and $\beta$ are the standard Dirac matrices$^{10}$

$$ {\gros \alpha}= \pmatrix{ 0&  {\gros\sigma}\cr
                           {\gros\sigma}&   0 }, \qquad  \beta = \pmatrix{1& 0\cr
               0& 1}, \eq{2}  $$

\noindent where the 1's and 0's stand, respectively, for $2\times2$ unit and zero matrices and the $\gros\sigma$ is the standard vector composed by the three Pauli matrices ${\gros \sigma}=(\sigma_x, \sigma_y, \sigma_z)$. Since the Hamiltonian (1)  is invariant under rotations, the solutions of the problem can be written in the form

$$
\psi(r,\theta,\phi) = {1\over r}\left( \matrix{F(r){\cal Y}_{jm}(\theta, \phi)\cr \cr iG(r){\cal Y}'_{jm}(\theta,\phi)}\right). \eq{3} $$

\noindent Where ${\cal Y}_{jm}$ and ${\cal Y}'_{jm}$ are  spinor spherical
harmonics of opposed parity. Parity is a good quantum number in the problem because the Coulomb potential is invariant under reflections; parity goes as $(-1)^l$ and, according
to the triangle's rule of additon of momenta,  the orbital angular
momentum $l$  is given by $l=j\pm {1\over2}$. But, instead of working
directly with parity, we prefer to introduce the quantum number $\epsilon $
defined by

$$ \epsilon =\cases{1 & If $ l=j + {1\over 2},$\cr
\cr
-1 & If $ l= j- {1\over 2}$.}
\eq{4} $$

\noindent Thus $l=j+{\epsilon\over 2}$  in all cases;  we also define $ l'=j
- {\epsilon \over 2}$. Accordingly, the spherical spinor ${\cal Y}_{jm}$
depends on $l$ whereas the spherical spinor ${\cal Y}'_{jm}$, which has an
opposite parity, depends on $l'$. Writing the solutions in the form (3) completely solves the angular part of the problem. \Par

Let us now address the radial part of the problem; we are interested in its  bound states, then the quantity $k\equiv\sqrt{m^2 - E^2}$ is  positive definite; furthermore, let us  define

$$
 \zeta\equiv Ze^2,\quad \tau_j\equiv \epsilon(j+{1\over 2}), \quad \nu \equiv \sqrt{m-E\over m+ E},
\eq{5} $$

\noindent then,  we can write the differential equations for the radial part of
the problem, in terms of the dimensionless variable $\rho = kr$,  as

$$
\left(-{d\over d\rho} + {\tau_j\over \rho}\right)G(\rho) = \left( -\nu
+{\zeta\over \rho}\right)F(\rho), \eq{6} $$

\noindent and

$$
 \left(+{d\over d\rho} +
{\tau_j\over\rho}\right)F(\rho) = \left( \nu^{-1} +
{\zeta\over\rho}\right)G(\rho); \eq{7} $$

\noindent these equations are to be regarded as the initial formulation of our problem. \Par

The first thing we want to do is to show  that Eqs.\ (6) and (7) can be rewritten using a set of three operators whose commutation relations define a SU(2) algebra.   To this end, let us first introduce the  new variable $x$ through the relation (but please notice that this change is not required for any of the conclusions that follow, see the second remark below after Eq.\ (15)) 

$$ \rho=e^x,  \eq{8}$$

\noindent so $x$ is defined in the open interval $(-\infty, \infty)$ and redefine the radial functions  $F(\rho)$ and $G(\rho)$, introduced in equations (6) and (7), in the form

$$
F(\rho(x))  = \sqrt{m + E}\,\left[\psi_{-}(x) + \psi_{+}(x)\right], \eq{9} $$

$$G(\rho(x))  = \sqrt{m - E}\,\left[\psi_{-}(x) - \psi_{+}(x) \right]. \eq{10}
$$

In terms of the new functions $\psi_+(x)$ and $\psi_-(x)$, we thus arrive to the following  set of equations  for our problem

$$
\left[{d\over dx} + e^x - {\zeta E\over\sqrt{m^2 - E^2}}
\right]\psi_+(x)  =  \left({\zeta m\over \sqrt{m^2 - E^2}} - \tau_j
\right)\psi_-(x), \eq{11}
$$

\noindent and

$$ -\left[{d\over dx} -e^x + {\zeta
E\over\sqrt{m^2 - E^2}} \right]\psi_-(x)  =  \left({\zeta m\over
\sqrt{m^2 - E^2}} + \tau_j\right)\psi_+(x). \eq{12} $$

\noindent This first-order system can be uncoupled  multiplying  by the left the first equation (Eq.\ (11)) times the operators that appear between
square brackets in the second equation and, viceversa, by  multiplying the
second equation (Eq.\ (12)) times the operators that appear (between square brackets) in the first one.
This procedure gives us the second order  system

$$
\left[{d^2\over dx^2} +  2\mu e^x - e^{2x}-
{1\over 4} \right]\psi_+(x)  =  \left( {\tau_j}^2 -\zeta^2 - {1\over
4}\right)\psi_+(x), \eq{13} $$

\noindent and

$$ \left[{d^2\over dx^2}+
2\left(\mu -1\right)e^x  - e^{2x} -{1\over 4} \right]\psi_-(x)  =
\left(\tau_j^2 -\zeta^2 - {1\over 4}\right)\psi_-(x), \eq{14} $$

\noindent where we have defined

$$
\mu \equiv {\zeta E\over \sqrt{m^2 - E^2 }} + 1. \eq{15}
$$

At this point there are several remarks that need  to be done. First,  in the next
section  the seemingly odd term $-1/4$ in Eqs.\ (13) and (14) is shown to be necessary to construct the
algebra. Second,  as we said before, the change of variable from $\rho$ to $x$ is not really necessary for any of the calculations in the article, however, we prefer to work in the $x$ rather than in the $\rho$ variable because this choice simplifies the appearance of some of the equations and, mainly, because it makes the inner product introduced in section III (in Eq.\ 33) look  more familiar; but to make contact with the usual description we sometimes, at our convenience, revert to the variable $\rho$. As a third remark, notice that we can regard equations (13) and (14) as two eigenvalue equations where the common eigenvalue $\omega$ is given by

$$
\omega = \tau_j^2 - \zeta^2 -{1\over 4} = j(j+ 1) - \zeta^2; \eq{16}
$$

\noindent as it  is obvious, we do not need to calculate $\omega$ because it follows directly from the radial symmetry of the problem and from the intensity of the interaction which is needed to set the scale.
  The fourth remark we want to do is that, as the minimum value of $j$ is
$1/2$, then $\omega \geq 0$ for at least $Z= 1,2,\cdots $ up to 118; for a discussion of the significance of this number see Ref.\ 4, page 236. \Par

\noindent{\bf III. An  operator algebra for the Dirac hydrogen atom}\Par

As the main purpose of this article is the construction of  non unitary representations of the SU(2) algebra for the Dirac hydrogen atom,$^{3}$ let us first introduce the operator

$$
\Omega_3 \equiv -i{\partial \over \partial \xi}, \eq{17} $$

\noindent depending exclusively on a new variable $\xi$, which is essentially an extra phase as must be clear in what follows,  and the two  operators

$$
\Omega_{\pm} \equiv e^{\pm i\xi}\left({\partial \over \partial x} \mp
e^x \mp i{\partial\over \partial \xi} + {1\over 2}\right). \eq{18} $$

\noindent which depend both on $\xi$ and on the transformed `radial' variable $x$. These three operators satisfy the following algebraic relations

$$
\left[\Omega_3 ,\Omega_{\pm}\right] = \pm \Omega_{\pm}, \eq{19}
$$
\noindent and

$$
\left[ \Omega_+, \Omega_-\right]=  2\Omega_3. \eq{20}
$$
\noindent  We can alternatively define the two operators $\Omega_1$ and $\Omega_2$ as

$$ \Omega_1= {1\over 2} \left(\Omega_+ + \Omega_-\right), \qquad \Omega_2= {1\over 2i} \left(\Omega_+ - \Omega_-\right), \eq{21} $$

\noindent in terms of which the algebraic properties (19) and (20) read

$$ [\Omega_i, \Omega_j]= i \epsilon_{ijk} \Omega_k, \quad i,j,k =1,2,3. \eq{22}$$

\noindent It is now clear that the conmutation relations (19) and (20) ---or just (22) --- correspond to an SU(2) algebra.$^{1,3,8}$ To complete the discussion, we  also need to introduce the Casimir operator of the algebra;  let us  consider the operator

$$
\Omega^2 = \Omega^2_1 + \Omega_2^2 +\Omega_3^2, \eq{23}
$$

\noindent we can easily show that (23) is indeed a Casimir for the algebra (19) and (20) (or (22))

$$
\left[ \Omega^2,\Omega_{i}\right] = 0.\quad\hbox{for $i= 1,2,3$}. \eq{24}
$$

\noindent Here, as in the angular momentum case, we can regard $\Omega^2$ as the square of $\gros{\Omega}= \Omega_1 {\bf\hat i} +  \Omega_2 {\bf\hat j} + \Omega_3 {\bf \hat k}$.  To obtain the explicit expression of $ \Omega^2$, it is better to calculate first the product $\Omega_-\Omega_+$:

$$
\Omega_-\Omega_+ = {\partial^2\over \partial x^2} - e^{2x} -2ie^x{\partial\
\over \partial\xi} + i {\partial \over \partial \xi} +
{\partial^2\over\partial\xi^2}- {1\over 4} \eq{25}
$$

\noindent and obtain the Casimir
operator from the relationship  $\Omega^2 = \Omega_-\Omega_+ + \Omega_3(\Omega_3-1)$. We remark that we do not have  a linear term in $\partial/\partial x$ in equation (22) because we choose the last constant in equation (18) precisely as $1/2$.  We then easily conclude  that the Casimir operator is given  explicitly by

$$
\Omega^2 = {\partial^2\over \partial x^2} - e^{2x} -2i e^x
{\partial\over\partial \xi}- {1\over 4}. \eq{26}
$$

Although we are not restricted to a compact set of parameters anymore due to the presence of the variable $x$, in analogy with the conventions used for the {\it spherical harmonics} $Y_l^m(\theta,\phi)$, we label the simultaneous eigenfunctions of $\Omega^2$ and $\Omega_3$ as $V_\omega^\mu(x,\alpha)$; where the numbers $\omega$ are the eigenvalues of $\Omega^2$ and the numbers $\mu$ are the eigenvalues of $\Omega_3$. Therefore, we write

$$
\Omega_3 V_\omega^\mu (x,\xi) = \mu V_\omega^\mu (x,\xi) \eq{27}
$$

\noindent and

$$\Omega^2V_\omega^\mu (x,\xi) = \omega V_\omega^\mu(x,\xi); \eq{28}
$$

\noindent we thus have

$$
V_\omega^\mu(x,\xi) = e^{i\mu \xi}{\cal P}_\omega^\mu (x). \eq{29}
$$

\noindent Where, again, we have used the angular momentum analogy to write the $x$ functions as
${\cal P}_\omega^\mu (x)$, resembling the usual notation for the Legendre polinomials $P_l^m(\theta)$. In this equation it becomes clear the role of $\xi$ as just an extra phase. \Par

Notice that Eqs.\ (26) and (28) serve to establish the connection between the introduced operators of the SU(2) algebra and the radial part of the hydrogen atom problem,  since applying the Casimir operator $\Omega^2$ to $V_\omega^\mu (x,\xi)$ or to $V_\omega^{\mu-1}(x,\xi)$, reproduces Eqs.\ (13) and (14). Furthermore, comparison with these same equations  tell us  that

$$
\psi_+(x) ={\cal P}_\omega^\mu(x), \eq{30}
$$

$$ \psi_-(x) ={\cal P}_\omega^{\mu -1}(x). \eq{31}
$$

\noindent The  operators $\Omega_{\pm}$ play the role of ladder operators for the problem;  they move along the set of eigenfunctions changing the eigenvalue $\mu$ to the eigenvalue $\mu\pm 1$, in a completely analogous way to the the case of the angular momentum algebra,

$$
\Omega_{\pm}V_\omega^\mu (x,\xi) \propto V_\omega^{\mu \pm 1}(x,\xi), \eq{32}
$$
\noindent the required proportionality constants are evaluated in section V (Eq.\ (52)).\Par

\noindent{\bf IV. The inner product and related properties of the operators}\Par

To establish the properties of the $\Omega$-operators and to construct the representations of the SU(2) algebra they span, we need  a properly defined inner product. Here we cannot longer rely on the angular momentum analogy, since two of the parameters of the algebra  we purport to construct are defined over the non-compact interval $ (-\infty,\infty)$, making it completely different from the angular momentum.  This situation implies that not all the generators of the algebra are to be Hermitian (indeed, $\Omega_3$ is found to be Hermitian but $\Omega_1$ and $\Omega_2$ are are found to be anti-Hermitian) and, as a
consequence, that the corresponding group could not be unitary. \Par

Let us consider the operation

$$
(\phi,\psi) = \int_0^{2\pi}{d\xi\over 2\pi}\int_{-\infty}^\infty
\phi^{*}(\xi,x)\psi(\xi,x)\,dx.  \eq{33}
$$

\noindent It is easy to show that Eq.\ (33) defines an inner product, since, as it is not difficult to prove, it satisfies the three basic properties:\Par

\noindent \item {\it i}) If $\psi(x)=0,$ then $(\psi,\psi)=0$.\par

\noindent \item {\it ii}) If $\psi (x)\neq 0$, then $(\psi,\psi)\geq 0$.\par

\noindent \item {\it iii})  If $c$ is any complex number, then $(\psi, c\phi)= c(\psi,\phi)$ and $(c\psi, \phi)= c^*(\psi,\phi)$. \Par

 To study the behavior of the generators of the algebra in terms of the inner product (33), we consider a certain function $\psi(x,\xi)$ associated to a certain fixed value $\mu_0$. We define the elements of the associated Hilbert space $H_{\mu_0}$ with functions of the form

$$   \psi(x, \xi)=e^{i(\mu_0+m-n)\xi}{\cal F}(x), \eq{34} $$

\noindent where $m$ and $n$ are integer numbers and ${\cal F}(x)$ is a well behaved function depending only on $x$. As we exhibit in section V, this is actually the general form of the functions inhabiting the Hilbert space of our problem.
\Par

With the help of the inner product (33), we can establish the properties of the $\Omega_a$ ($a=1, 2,  3$) operators. Let us consider first $\Omega_3$, in this case the important product is

$$ (\psi', \Omega_3 \psi)=\,\delta_{m'-n',m-n}\,(\mu_0+m-n)\int_{-\infty}^{\infty} {\cal F}^{ *}(x) {\cal F}(x)\, dx; \eq{35} $$

\noindent  this last equation, due to the presence of the Kronecker delta, can be written as

$$ \int_{0}^{2\pi} {d\xi\over 2\pi} \int_{-\infty}^{\infty} \left[ i{\partial\over \partial\xi}
e^{-i(\mu_0+m'-n')\xi} \right]e^{-i(\mu_0+m-n)\xi}{\cal F}^{*}(x){\cal F}(x)\, dx, \eq{36} $$

\noindent and this is precisely $(\Omega_3\psi', \psi)$. We have in this way proved that $\Omega_3$ is an Hermitian operator:

$$ \Omega_3^{\dag} =\Omega_3. \eq{37} $$

To study the operators $\Omega_\pm$, let us first consider $\Omega_+$ and evaluate the product

$$ \eqalign{(\psi', \Omega_+\psi) = &\int_{0}^{2\pi}{d\xi \over 2\pi}\int_{\infty}^{\infty}e^{-i(\mu_0+m'-n')\xi}
{\cal F}^{*}(x)\, dx \cr
 & e^{i\xi}\left( {\partial \over \partial x} -
e^x - i{\partial\over \partial \xi} + {1\over 2} \right)e^{i(\mu_0+m-n)\xi}{\cal F}(x)\, dx.
} \eq{38} $$

\noindent To analyse this integral, it is simpler to split it in two parts. Let us consider first the $e^{i\xi}(-\partial/\partial\xi + 1/2)$ part; its contribution to the inner product (38) is

$$ \left(\mu_0+m-n+{1\over 2}\right)\,\delta_{m'-n',m-n+1}\int_{-\infty}^{\infty}{\cal F}^{*}(x){\cal F}(x)\, dx, \eq{39}  $$

\noindent this expression can be written as

$$ \eqalign{ -\int_{0}^{2\pi}{d\xi \over 2\pi} \left[e^{-i\xi}\left(i{\partial\over \partial\xi}+{1\over 2} \right) e^{i(\mu_0+m'-n')\xi} {\cal F}(x)  \right]^\dagger \cr
  e^{i(\mu_0+m-n)\xi} &{\cal F}(x) \, dx.} \eq{40} $$

\noindent We now take care  of the contribution of the term $e^{i\xi}(\partial/\partial\xi -e^{x})$. After a partial integration this contribution becomes

$$  \delta_{m'-n'-1,m-n}\int_{-\infty}^{\infty}\left[-\left({\partial\over \partial x}+e^x \right) {\cal F}(x) \right]^\dagger{\cal F}(x)\, dx. \eq{41} $$

\noindent Taking together the two previous results, it is easy to see that the operator complies
with $\Omega_+^\dagger= -\Omega_-$. A similar calculation  establish the analogous property $\Omega_-^\dagger=-\Omega_+$. Therefore, we have established that

$$ \Omega_\pm^\dagger= -\Omega\mp. \eq{42}$$

We  have proved that not all the operators are  Hermitian thence  arriving to the anticipated results: First, we showed that the operator $\Omega_3$ is indeed Hermitian, as a consequence we expect the range of the only parameter it depends on, $\xi$, to be compact; this is certainly the case since $\xi\in [0, 2\pi]$, a compact set. The other two operators $\Omega_1$ and $\Omega_2$ depend upon $x$, a variable defined over the non compact set $(-\infty, \infty)$, but here such operators are anti-Hermitian  

$$\Omega_a^\dagger=-\Omega_a,\qquad a=1,2 \eq{43}$$

\noindent as it is easy to show from Eqs.\ (21) and (42). \Par

\noindent{\bf V. Representations of the SU(2) algebra (radial eigenfunctions of the problem).  }\Par

With the inner product defined in the previous section, we are in the position of introducing a complete orthogonal basis of simultaneous eigenfunctions for $\Omega^2$ and $\Omega_3$ which accordingly must carry a representation of the algebra---and, besides, they solve our radial eigenvalue problem (Eqs.\ (6) and (7)). We have  decided to choose the conmuting operators in similar fashion to the standard SU(2) case. Let us define then

$$  V_\omega^\mu (x,\xi)\equiv |\omega\,\mu>, \eq{44}  $$

\noindent where the kets $|\omega\,\mu>$ are both assumed orthogonal and normalized respect the inner product (33)

$$  <\omega'\,\mu' | \omega\,\mu> = \delta_{\omega ,\omega'}\delta_{\mu,\mu'}. \eq{45}  $$

Now, as $\Omega_1$ and $\Omega_2$ are not Hermitian, the Casimir operator $\Omega^2$ defined in (23) is not necessarily positive definite, we can nevertheless introduce a positive definite operator as

$$ {\gros\Omega}^\dagger\cdot{\gros\Omega}=-\Omega_1^2-\Omega_2^2+\Omega_3^2=2\Omega_3^2-\Omega^2; \eq{46} $$

\noindent the positivity of this operator allows us to show that

$$ 2\mu^2\ge \omega, \eq{47}$$

\noindent that is, $|\mu|$ is bounded by below. As a consequence, there must exist a minimum value for $|\mu|$, let us say $\lambda\equiv |\mu|_{\
\hbox{\rm min}}$. We also know the that the ket $|\omega \,\lambda> $ is annihilated by $\Omega_-$ or, equivalently, that $\Omega_+\Omega_- |\omega\, \lambda>=0$; so

$$  \omega-\lambda^2-\lambda=0 \qquad \hbox{ or } \qquad \omega= \lambda(\lambda-1). \eq{48} $$

\noindent Given equation (47), let us introduce a slight change in notation, writing $\lambda$ instead of $\omega$ in the eigenfunctions ${\cal P}^{\mu}_\omega(x)$ of Eq.\ (29), that is, ${\cal P}^{\mu}_\omega(x)$ is to be replaced by  ${\cal P}^{\mu}_\lambda(x)$.  Furthermore, as $\lambda$ has to be positive and since $\omega=\tau_j^2-\zeta^2-{1/ 4}$, we find
that the minimum $|\mu|$-value is

$$|\mu|_{\hbox{min}}\equiv \lambda=s+{1\over 2},  \eq{49}  $$

\noindent where we have defined

$$ s\equiv +\sqrt{\tau_j^2-\zeta^2}=\sqrt{\left(j+{1/2}\right)^2-\zeta^2}. \eq{50}  $$

\noindent Notice that $s$ is a real quantity for $Z=1, 2, 3, \dots, 137$.\Par

 However, the most important conclusion we can draw from Eqs.\ (49) and (50), is that $\lambda$ no longer has to be an integer or half-integer number, as necessarily happens  in the unitary compact angular momentum case. We may see that this result is a direct consequence of both $\Omega_1$ and $\Omega_2$ not being Hermitian operators. So the operators introduced for the SU(2) algebra associated with the Dirac hydrogen atom lead  naturally to non unitary representations labeled by real numbers, or, at least, not necessarily integer nor half-integer ones. \Par

For constructing the functions comprising the representations, let us introduce the constants $C_\mu^{\pm}$ as follows (compare with Eq.\ (32))

$$ \Omega_{\pm} |\omega\,\mu>= C_\mu^{\pm}|\omega\,\mu\pm 1>; \eq{51} $$

\noindent these constants can be explicitly evaluated from $<\omega\,\mu|\Omega_+\Omega_-|\omega\,\mu>=C_\mu^{-}C_{\mu-1}^{+}$ and from $(C_{\mu-1}^{+})^*=-C_\mu^{-}$; if we further assume these constants to be real, we easily get

$$   C_\mu^{\pm}=\pm\sqrt{\mu(\mu\pm1)-\lambda(\lambda-1)}, \eq{52} $$

\noindent a result with a slightly different form than in the analogous angular momentum case.$^{4,8}$ \Par

 The ground state of the hydrogen atom, can be obtained from the equation $\Omega_-|\lambda\,\lambda>=0$  for the positive set of eigenvalues. Such equation becomes 

$$ e^{-i\xi}\left(  {\partial\over\partial x} +e^x-\lambda+{1\over 2} \right) e^{i\lambda\xi}{\cal P}^{\lambda}_\lambda(x)=0. \eq{53} $$

\noindent whose solution is

$$ {\cal P}^{\lambda}_\lambda(x)=d_\lambda\, e^{sx}\exp(-e^x)=d_\lambda \,e^{(\lambda-1/2)x}\exp(-e^x), \eq{54} $$

\noindent where

$$ d_\lambda\equiv{ 2^{(\lambda-1/2)}\over \sqrt{\Gamma(2\lambda-1)} } \eq{55}$$

\noindent is a normalization constant and $\Gamma(y)$ stands for the Euler-gamma function. Since $\lambda$ is the lowest eigenvalue, in this instance we should have

$$ \psi_+(x)={\cal P}^{\lambda}_\lambda(x), \eq{56} $$

\noindent and

$$ \psi_-(x) = 0. \eq{57}  $$

In terms of the radial variable $\rho$, we can see  from Eqs.\ (9) and (10) that the ground state solutions  are

$$ F(\rho)=\sqrt{(m+E)\over 2m(\lambda-1/2) } \rho^s e^{-\rho}, \eq{58}$$

\noindent and

$$ G(\rho)= -\sqrt{(m-E)\over 2m(\lambda-1/2) } \rho^s e^{-\rho}. \eq{59} $$

\noindent normalized in the sense

$$ \int_0^{\infty} \left( |F(\rho)|^2 + |G(\rho)|^2 \right) \, d\rho =1. \eq{60}  $$

We can also see that in the case of negative eigenvalues the solutions behave as $\sim \rho^s e^\rho $, giving divergent behavior as $\rho\to \infty$. This behavior makes the negative energy solutions not square integrable; therefore, we have to discard them if we want to describe  physically realizable states.$^{2,3,10}$ \Par

The excited states can be  obtained, as we do in the appendix,  by applying succesively $\Omega_+$ to the ground state $ |\lambda\, \lambda>$, as it is customarily done for the spherical harmonics. The result involves the functions 
${\cal P}^{\mu}_\lambda(x)$ introduced by Eqs.\ (29) and (34). The ${\cal P}^{\mu}_\lambda(x)$ are polynomials multiplied by the weight factor $W(\rho)=\rho^{\lambda-1/2}e^{-\rho}$. This weight factor assures that the behavior of the big and the small components of the spinor are regular both at the origin as well as $\rho\to\infty$. As illustration of the solutions discussed here, we quote in the Appendix the first few cases of $ {\cal P}^{\lambda+p}_\lambda(x)$, for $p=0,\dots, 5$, there we also plot these polynomial part of the functions (\ie\ we plot ${\cal P}^{\mu}_\lambda(x)$ without the weight factors) for the first 5 excited energy levels. Notice 
that the polynomial part in the eigenfunction  ${\cal P}^{\lambda+p}_\lambda(x)$ is always of order $p$.\Par

The matrix representations of the  $\Omega_a$, $a=1,2,3$, for $\lambda=s+1/2+p$, $p=0,1,2,3,\dots$, are constructed from equations (27), (28), (51) and (52). In contrast with the standard SU(2) case, here the representations are non Hermitian and infinite dimensional for each $\lambda$, excepting for $\Omega_3$. The matrix elements of the operator $\Omega_3$, including the negative eigenvalue series, are given by

$$ <\omega\mu|\Omega_3|\omega\mu'>=\mu\, \delta_{\mu\mu'}, \eq{61} $$

\noindent where $\mu=\pm(\lambda+p)$, $p=0,1,2,\dots$; therefore, its trace vanishes and the determinant of any element   of the corresponding group with  the form $\exp(i\Omega_3)\xi$, is always 1. \Par

For the other two operators the only non-vanishing matrix elements are

$$ <\omega\mu|\Omega_1|\omega\mu\pm1>=\mp{1\over 2} \sqrt{\mu(\mu\pm1)-\lambda(\lambda-1)}, \eq{62} $$

$$ <\omega\mu|\Omega_2|\omega\mu\pm1>=
-{i\over 2} \sqrt{\mu(\mu\pm1)-\lambda(\lambda-1)}, \eq{63} $$

\noindent This means again that the trace of both $\Omega_1$ and $\Omega_2$ vanish. Notice also that the determinant of a group element generated by $\Omega_1$ or $\Omega_2$ is 1 only for a purely imaginary parameter. \Par

\noindent{\bf VI. The energy spectrum.}\Par

We can now evaluate the bound energy spectrum for the problem. As we mentioned in section V, the bound state energy spectrum comprises only  the positive series eigen\-values.$^{2,10}$ Let us first  express the energy in terms of the eigenvalue $\mu$ from Eq.\ (15)

$$ E=m\left[1 + {\zeta^2 \over (\mu -1/2)^2}\right]^{-1/2}. \eq{64} $$

\noindent then, for the case of positive eigenvalues ---the only ones with physically appropriate eigenfunctions--- we have that $\mu=\lambda+p$, where $p$ is an non negative integer $p=0,1,2,\dots$,  or, equivalently,  that $\mu-1/2=s +p$. This gives precisely the energy spectrum of the relativistic hydrogen atom. To rewrite our result in a more familiar form, we need only to define the principal quantum number $n$ and the auxiliary quantity $\epsilon_j$ as follows

$$ n= j + {1\over 2} +p, \eq{65}$$

$$ \epsilon_j= n-s-p = j + {1\over 2}-s; \eq{66} $$

\noindent we then finally conclude that $\mu-1/2=s+p=n-\epsilon_j$,  which gives precisely the well-known energy spectrum.$^{10}$ \Par

\noindent {\bf VII. Concluding remarks.} \Par

In summary, we have constructed an SU(2) algebra for the  hydrogen atom in the Dirac formulation, introducing the Hermitian operator  $\Omega_3$ and the anti-Hermitian operators $\Omega_1$ and $\Omega_2$.  The result of all of this, is that the representations are labelled by numbers $\lambda$ wich are neither integers nor half-integers as in the case of the  more familiar unitary representations. Nevertheless, the algebra introduced predicts precisely the energy eigenvalues and eigenfunctions of the Dirac hydrogen atom.  One of the most noteworthy features of the representations reported here is the mixing of a  spinorial angular momentum character, implying an equally spaced spectrum, with the energy requirements of the problem---requiring a differently spaced spectrum; the interplay of these two spectral requirements is basically reflected in the fact that the eigenvalues associated with Eqs.\ (13) and (14) follows from both the generic radial symmetry  and the specific features of the interaction in the Dirac equation (1).  From Eq.\ (15) it also follows that in the limit of vanishing interaction,\ie\ $\zeta\to 0$, our representations collapse and, in this special case, $\mu=1$ always. Such behavior is precisely as expected because there is no longer any restriction over the eigenvalues and thus the spectrum becomes continuous, corresponding to a free Dirac particle. The operator algebra we introduced allows an essentially algebraic solutiond of Dirac hydrogen atom which may have various applications.$^{3,4,6}$

   It is to be noted also the possible connections that our formulation may have with systems with hidden supersymmetric properties,$^{13\hbox{--}16}$  as we will discuss  in a forthcoming article. The energy spectrum of the problem has some peculiarities which also appear in the spectrum of a Dirac oscillator;$^{14\hbox{--}15,17}$ in particular the equally spaced energy solutios for $\psi_+(x)$ and $\psi_-(x)$ resemble, respectively, the behavior of big and the small components of the aforementioned system. As a result of this resemblance, we are studying the hidden supersymmetric properties and the superconformal algebra, in the sense of Refs.\ 15 and 18, associated with this problem. The similitude of Eqs.\ (13) and (14) with the corresponding ones for a Morse oscillator should be also noticed.$^6$ \Par

 It is worth pinpointing that we are forced to introduce the new variable $\xi$ in order to define the algebra; in terms of the solutions of the Dirac equation, $\xi$ just plays the role of a phase.
To exemplify, when we perform a ``rotation'' using $\Omega_3$, the big component changes from $F(x)\propto [\psi_-(x) + \psi_+(x)] $ to $F(x)\propto [e^{i(\mu-1)\xi}\psi_-(x) + e^{i\mu\xi}\psi_+(x)]= e^{i\mu\xi}[e^{-i\xi \psi_-(x) + \psi_+(x)]}$. The phase $e^{i\mu\xi}$ does not play any observable role, but the term $e^{-i\xi}$ changes the relative phase between the $\psi_+(x)$ and the $\psi_-(x)$ components of the eigenfunction and, in consequence, changes the radial function $F(\rho)$ itself although the energy spectrum is still invariant under such transformation; this is a consequence of the fact that  $|\omega\,\mu>$ and $\Omega_3|\omega\,\mu>$ both correspond to the same eigenvalue $\mu$. In a way, this resembles what happens when there are superselection rules in a system.$^{19\hbox{--}21}$ \Par

 To finalize, let us comment that it is  not widely known that the radial eigenfunctions of the Dirac hydrogen atom can be expressed in terms of generalized associated Laguerre polynomials, as was realized by Davis a long time ago.$^{7,11}$ These polynomials, which are a generalization to  non integral indices of the usual associated Laguerre polynomials, are defined as$^{7}$ \Par

$$ {\cal L}_p^{\alpha}(x)={\Gamma(\alpha+p+1)\over n! \Gamma(\alpha + 1)} {}_1F_1(-p;\alpha+1;x) \eq{67} $$

\noindent where  $\Gamma(x)$ is again the Euler gamma function, $p$ is a positive integer, and the ${}_1F_1(-p,\alpha+ 1; x)$ stands for the confluent hypergeometric function ---having one of their arguments negative, the hypergeometric function reduces to a polynomial.$^{12}$  The polynomial representation of the radial eigenfunctions introduced here is related to that used by Davis in Eqs.\ (A8) and (A9) of the Appendix. \Par

\noindent{\bf Acknowledgements.}\Par

\noindent This work has benefitted of the insightful comments of L.\ F.\ Urrutia, R.\ Jauregui, and H.\ N.\ N\'u\~nez-Y\'epez and has been partially supported by CONACyT (grant 1343P-E9607). ALSB acknowledges the help of F.\ C.\ Minina, G.\ Tigga, U.\ Kim, Ch.\ Ujaya, B.\ Caro,  M.\ X'Sac, M.\ Osita, Ch.\ Dochi, Ch.\ Mec and F.\ C.\ Bonito. Last but not least this paper is dedicated to the memory of Q.\ Motita, N.\ Kuro, M.\ Mina, M.\ Tlahui and M.\ Miztli. \Par

\noindent{\bf Appendix} \Par

 The  purpose of this Appendix is to illustrate the behavior of the first few polynomials associated to the radial eigenfunction of the hydrogen atom in the base $|\lambda \, \,\mu>$ and to relate this description to the old but little known results of Davis.$^{7,11}$ We explicitly calculate  the first 6 functions of the positive eigenvalue series, such functions are always of the form the weight factor $W(\rho)$ times a polynomial; the weight factor is $W(\rho)=\rho^{s} e^{-\rho}$.   The polynomials are  plotted in Figure 1. The radial eigenfunctions  are essentially  generated from the basic relationship $\Omega_+|\omega\, \mu>= C_\mu^{+}|\omega\, \mu + 1>$. In  the  $\rho$ variable, the first equation in the  series   can be written as

$$ e^{+ i \xi}\left(\rho {\partial \over \partial \rho} - \rho - i {\partial \over \partial \xi} + {1\over 2}  \right)e^{ i \mu \xi} {\cal P}^{\lambda}_\lambda(\rho)=C_\mu^{+} e^{i(\mu+1)\xi}{\cal P}^{\lambda+1}_\lambda(\rho) \eq{A1}$$

\noindent which is just the first term in the whole ascending series ${\cal P}^{\lambda+p}_\lambda(\rho)= \Omega_+^{p}{\cal P}^{\lambda}_\lambda(\rho)$ used to recursively calculate  (A2--A7). Note that the polynomial part of the function ${\cal P}^{\lambda+p}_\lambda(x)$ is always of the form $ \sum_{i=0}^p C_i(\lambda) \rho^i$, where $C_i(\lambda)$ is also an order $(p-i)$ polynomial in $\lambda$.\Par

Starting with $\mu=\lambda$, the first few functions in the positive series are then given by

$$ \eqalignno{
          {\bf {\cal P}^{\lambda}_\lambda(\rho)}=&  W(\rho), &(A2)\qquad \cr
          {\bf {\cal P}^{\lambda+1}_\lambda(\rho)}=&  \sqrt{2\over \lambda} (\lambda-\rho)W(\rho),
&(A3)\qquad \cr
{\bf {\cal P}^{\lambda+2}_\lambda(\rho)} =& {      W(\rho)        \over \sqrt{2\lambda(\lambda + 1/2)}}[2\rho^2 -(2\lambda + 1)(2\rho -\lambda)],&(A4)\qquad \cr
{\bf {\cal P}^{\lambda+3}_\lambda(\rho)}=& {     W(\rho)         \over \sqrt{3\lambda(\lambda + 1/2)(\lambda +1)}}[-2\rho^3 + 6\rho^2 \,(\lambda + 1)- 3\rho\,(2\lambda^2 + 3\lambda + 1)\cr
+& \lambda\,(2\lambda^2 + 3\lambda + 1)], &A(5)\qquad \cr
{\bf {\cal P}^{\lambda+4}_\lambda(\rho)}=& \sqrt{2\over 3}{W(\rho)      \over \sqrt{\lambda(\lambda + 1/2)(\lambda +1)(\lambda + 3/2)}}[\rho^4 - 2\,(3 + 2\lambda)\rho^3 + 3\, (3 + 5\lambda + 2\lambda^2)\rho^2\cr
-& \,(3 + 11\lambda + 12\lambda^2 + 4\lambda^3)\,\rho + (\lambda^3 + 3\lambda^2 + {11\over 4}\lambda + {3\over 4}) \lambda ], &(A6)\qquad \cr
{\bf {\cal P}^{\lambda+5}_\lambda(\rho)} =& {   W(\rho)      \over \sqrt{15\lambda(\lambda + 1/2)(\lambda + 1)(\lambda + 3/2)(\lambda + 2)}}[-\rho^5 + (5\lambda + {29\over 2})\,\rho^4 \cr
-&\, ( 10\lambda^2 + 51\lambda + 54)\, \rho^3 + (10\lambda^3 + 66\lambda^2 + {235\over 2}\lambda + {123\over 2})\, \rho^2\cr
-& \, (5\lambda^4 + 37\lambda^3 + {319\over 4}\lambda^2 + {257\over 4}\lambda + {33\over 2})\,\rho +  \lambda^5 + {15\over 2}\lambda^4 + {65\over 4}\lambda^3\cr
+& {105\over 8}\lambda^2 + {27\over 8}\lambda ]. &(A7)\qquad } $$

The relationship to these polynomials to those used by Davis can be seen from Eqs.\ (13) and (14), putting $\rho=\exp x$ and introducing  the function $v(\rho)$ according to $\psi_+ \equiv \rho^s \exp(-\rho)\,v(\rho),$ and finally using ${\cal L}(\rho)\equiv v(\rho /2)$,
to  get

$$ \rho {d^2{\cal L}\over d\rho^2} + \left[ (2s+1)-\rho\right]  {d{\cal L}\over d\rho} +\left[{(s^2+\zeta^2-\tau_j^2)\over \rho} +(\mu -s -1/2)\right]{\cal L}= 0, \eqno(A8)\qquad$$

\noindent where $\mu=s + 1/2 + p$. This equation can be regarded as a generalization to non integer index of the usual associated Laguerre differential equation. The equation corresponding to $\psi_-$ can be obtained in analogous fashion. This equation reduces to the one used by Davis if and only if $s^2=\tau_j^2-\zeta^2,$ a result which just recovers the definition in Eq.\ (50). Now we can give the explicit relationship between our representation of the radial eigenfunctions to that used by  Davis  as

$$ \eqalign{
    \psi_+(\rho)=&{\cal P}^{s+1/2+p}_{s+1/2}(\rho)=\rho^s \exp(-\rho) {\cal L}^{2s}_{p}(2\rho), \cr
    \psi_-(\rho)=&{\cal P}^{s-1/2+p}_{s+1/2}(\rho)=\rho^s \exp(-\rho) {\cal L}^{2s}_{p-1}(2\rho)  }
     \eqno(A9)\qquad$$

\noindent  save  for normalization factors (unimportant for the point at hand)  where $s=\lambda-1/2$, and where the generalized associated Laguerre polynomials used by Davis are defined in Eq.\ (67). \Par

\vfill
\eject

\noindent{\bf References}\Par

\noindent 1. C.\ Itzykson and M.\ Naunberger, Rev.\ Mod.\ Phys.\ {\bf 38}, 1 (1966).\par

\noindent 2. Silvan S.\ Schweber, {\it An introduction to Relativistic Quantum Field Theory}, Harper and Row, New York, (1966).\par

\noindent 3.  R.\ P.\ Mart\'{\i}nez-y-Romero, J.\ Salda\~na-Vega and A.\ L.\ Salas-Brito, J.\ Phys.\ A: Math.\ Gen.\ {\bf 31}, L157 (1998). \par

\noindent 4. O. L. de Lange and R. E. Raab, {\it Operator Methods in Quantum Mechanics}, Clarendon Press, Oxford, (1991).\par

\noindent 5. L.\ Infeld and T. E. Hull, Rev. Mod. Phys. {\bf 23}, 21 (1951).\par

\noindent 6.  H.\ N.\ N\'u\~nez-Y\'epez, J.\ L.\ L\'opez, D.\ Navarrete and A.\ L.\ Salas-Brito, Int.\ J.\ Quantum Chem.\  {\bf 62}, 177 (1997). \par

\noindent 7. L.\ Davis Jr., Phys.\ Rev.\ {\bf 56} 186 (1939).\par

\noindent 8. A.\ B\"ohm, {\it Quantum Mechanics},  Springer-Verlag, New York, (1979), Ch.\ VI. \par

\noindent 9. A.\ L.\ Salas-Brito, R.\ P.\ Mart\'{\i}nez-y-Romero and H.\ N.\ N\'u\~nez-Y\'epez,  Int.\ J.\ Mod.\ Phys.\ A, {\bf 12}, 271 (1997).\par

\noindent 10. J.\ D.\ Bjorken and S.\ D.\ Drell  {\it Relativistic Quantum Mechanics}, Mac Graw-Hill, New York, (1964). \par

\noindent 11. H.\ A.\ Bethe and E.\ E.\ Salpeter, {\it Quantum Mechanics of One-- and Two--Electron Atoms}, Springer--Verlag, Berlin, (1957).\par

\noindent 12. J.\ B.\ Seaborn, {\it Hypergeometric Functions and their Applications}, Springer--Verlag, New York, (1991).\par

\noindent 13. J.\ Ben\'{\i}tez, R.\ P.\ Mart\'{\i}nez-y-Romero, H.\ N.\ N\'u\~nez-Y\'epez and A.\ L.\ Salas-Brito,  {Phys.\ Rev.\ Lett.\ } {\bf 64}, 1643 (1990);  {\bf 65}, 2085 (Erratum). \par

\noindent 14. M.\ Moreno and A.\ Zentella, J.\ Phys.\ A:  Math.\ Gen.\ {\bf 22}, L821 (1989). \par

\noindent 15. R.\  P.\ Mart\'{\i}nez-y-Romero  and A.\ L.\ Salas-Brito { J.\ Math.\ Phys.\ } {\bf 33} {1831}  (1992). \par

\noindent 16. R.\ P.\ Mart\'{\i}nez-y-Romero, M.\ Moreno, and A.\ Zentella, {Phys. Rev. D} {\bf 43} {2306} (1991). \par

\noindent 17. M.\ Moshinsky and J.\ Szcepaniak, J.\ Phys.\ A: Math.\ Gen.\ {\bf 22}, L821, (1989).\par

\noindent 18. R.\ Haag, V.\ T.\ Lopuszanski, and M.\ Sohnius, {Nucl. Phys. B} {\bf 88} {383} (1976). \par

\noindent 19.  R.\ P.\ Mart\'{\i}nez-y-Romero, H.\ N.\ N\'u\~nez-Y\'epez and A.\ L.\ Salas-Brito,   Rev.\ Mex.\ Fis.\   {\bf 35}, 617 (1989). \par

\noindent 20. R.\ P.\ Mart\'{\i}nez-y-Romero, H.\ N.\ N\'u\~nez-Y\'epez and A.\ L.\ Salas-Brito,  Phys.\ Lett.\ A {\bf 142} (1989) 318.\par

\noindent 21. C.\ Cisneros, R.\ P.\ Mart\'{\i}nez-y-Romero, H.\ N.\ N\'u\~nez-Y\'epez and A.\ L.\ Salas-Brito, Eur.\ J.\ Phys.\  {\bf 19}   237 (1998). \par

\vfill
\eject

\noindent {\bigfnt Figure Caption}\Par
\vskip 10 pt

\noindent Figure 1\Par
\vskip 7 pt
\noindent We show the graph of the first polynomials ${\cal P}^{\lambda+p}_\lambda(\rho)/W(\rho)$, for $p=1,2,3,4,5$, $Z=1$ and $j=1/2$. Notice the similarity of the behavior of all  polynomials near the origin.\Par

\vfill
\eject
\end
\voffset=-4cm
\input epsf.sty

\let\picnaturalsize=Y
\def\picsize{5 cm}
\def\picfilename{fig1.eps}
\ifx\nopictures Y\else{\ifx\epsfloaded Y\else\input epsf \fi
\global\let\epsfloaded=Y
\centerline{\ifx\picnaturalsize N\epsfxsize \picsize\fi \epsfbox{\picfilename}}}\fi

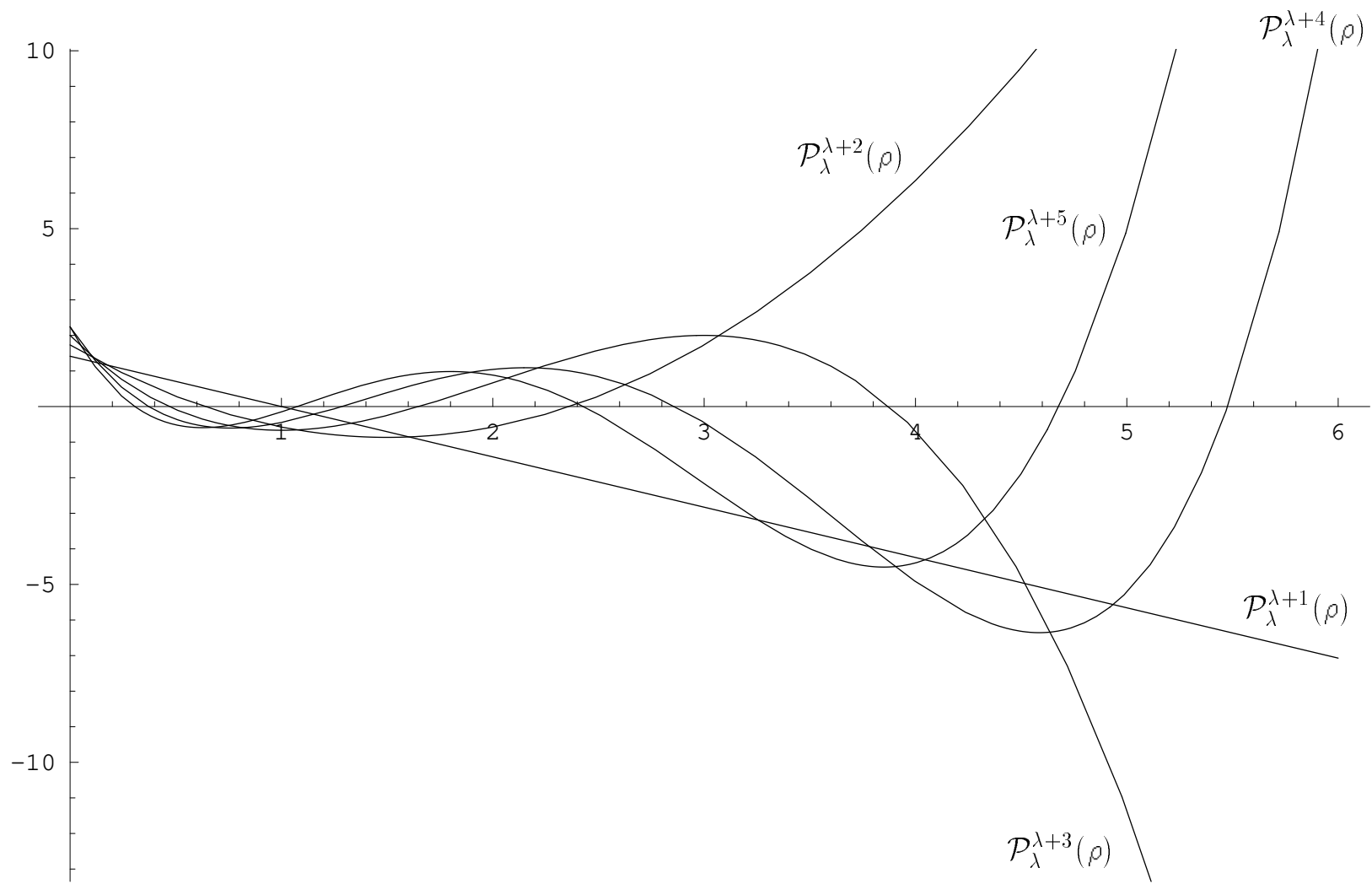
\bye